\documentclass[preprint,preprintnumbers,amsfonts,amsmath,amssymb,prd
,tightenlines]{revtex4}
\usepackage{graphicx}
\begin{document}
\preprint{FAU-TP3-09-01}
\title{Confining gauge fields $^{{\rm * }}$}
\author{F.Lenz $^{{\rm **}}$} 
\affiliation{Institute for Theoretical Physics III \\
University of Erlangen-N\"urnberg \\
Staudstrasse 7, 91058 Erlangen, Germany\\}
\date{September 17, 2009}
\begin{abstract}
By superposition of regular gauge instantons or merons, ensembles of gauge fields are constructed which describe the confining phase of SU(2) Yang-Mills theory.  Various properties of the Wilson loops, the gluon condensate and the topological susceptibility are found to be  in qualitative agreement with phenomenology or results of lattice calculations. Limitations in the application to the glueball spectrum and small size Wilson loops are discussed. 
\vskip 11cm \hspace{-.2cm}\small{*\, To appear in the proceedings of ``Shifmania, Crossing the Boundaries: Gauge Dynamics at Strong Coupling'' }\\ \small{** \,flenz@theorie3.physik.uni-erlangen.de}
\end{abstract} 
\maketitle
\section*{Introduction}
Despite convincing evidence for confinement in Yang-Mills theories from lattice gauge calculations and despite a host of analytical studies and conjectures, the intricacies of non-abelian gauge fields  have prevented an unambiguous identification of the  mechanism of confinement. In particular,  the gauge dependence of many of the mechanisms proposed constitutes a severe obstacle in reaching this goal. I will choose a different path towards understanding confinement. I will describe construction and analysis of ensembles of gauge fields  which exhibit  confinement and  account for  other facets of  the  dynamics of SU(2) Yang Mills theory. 
The studies to be presented  are based on the ideas that  instantons \cite{Belavin:1975fg} or merons \cite{deAlfaro:1976qz}, solutions of the classical Yang-Mills equations,  play a prominent role  also in the quantum theory \cite{Callan:1977gz, Callan:qs}. They are related to investigations which led to the  
instanton gas and liquid models  \cite{Shuryak:1981ff,Diakonov:1983hh} which successfully describe  a variety of  phenomena of strong interaction physics \cite{Schafer:1996wv,DIAK03}. These models  however miss confinement, the trademark of the dynamics of Yang-Mills fields. I will discuss an alternative  construction  of ensembles of gauge-fields  obtained by superposition of ``regular gauge'' instantons or of merons \cite{LNT48}. Although irrelevant for single instanton properties, the choice of either the regular or the singular gauge in the superposition of instantons  leads  to two different phases,   the gas or liquid phase of singular gauge  instantons with a well defined low density limit and  the  strongly correlated confining phase of regular gauge instantons (or merons).    
\section*{Gauge fields  of  instantons and merons} 
After appropriate choice of the coordinate system in color space  and
after regularization of  the singularity, the SU(2)  gauge field for a meron or an (regular gauge) instanton  in Lorenz gauge with the center  at the
origin,  is given by 
\begin{equation}
    \label{mer}
     a_{\mu}(x) = \xi\,\frac{\eta _{a\mu \nu } x_{\nu}}{x^{2} + \rho ^{2}}
\frac{\sigma^a}{2}\,,
\end{equation}
with the Pauli matrices $\sigma^a$, the 't Hooft tensor  $ \eta_{a \mu \nu}$ \cite{Schafer:1996wv},  and with $\xi=1$ denoting merons and $\xi=2$ instantons. 
 The size parameter $\rho$ controls the short distance  behavior of instanton and meron fields.  
For instantons, and for merons with vanishing  $\rho$, $a_{\mu}(x)$ is a solution of the  Euclidean classical  field equations  \cite{Belavin:1975fg,deAlfaro:1976qz}.

Important for the following are the infrared properties of the constituents.  The gauge fields of a single meron and  of a single  regular gauge  instanton  behave asymptotically as $1/x$ while the asymptotics of the field strengths, 
$$F^a_{\mu\nu}[A] = \partial_{\mu} A^a_{\nu}-\partial_{\nu} A^a_{\mu}+ \epsilon^{abc}A^b_{\mu}A^c_{\nu}\,,$$
is different for merons and instantons 
\begin{equation}
 x\to \infty\,,\quad F_{\xi=1}\to x^{-2}\,,\quad  F_{\xi=2}\to x^{-4}\,,
\label{fas}
\end{equation} 
due to the  cancellation of the abelian and non-abelian contributions to leading order for instantons. Therefore the  action density 
\begin{equation}
s(x) = \frac{1}{2}\, \mbox{tr}\, F_{\mu\nu}  F^{\mu\nu} 
\label{acde}
\end{equation}
behaves asymptotically as $x^{-4}$ and $x^{-8}$ for merons and instantons respectively and gives  rise to an infinite action for merons. 
The topological charge density is defined by 
\begin{equation}
\tilde s(x) = \frac{1}{2}\, \mbox{tr}\, F_{\mu\nu}  \tilde{F}^{\mu\nu} \,,
\label{tode}
\end{equation}
and yields a finite value for the topological charge 
\begin{equation}
\nu = \xi/2\,.
\label{toch}
\end{equation}
The fast asymptotic decrease of the field strength of an instanton is the basis for the alternate representation of instantons in ``singular gauge''
\begin{equation}
a^{\text{sing}}_{\mu} = \frac{2\rho^2}{x^2} \; \frac{\bar{\eta} _{a\mu \nu } x_{\nu}}{x^{2} + \rho ^{2}}\frac{\sigma^a}{2}\,,
\label{singin}
\end{equation}
where the gauge field decays asymptotically as   $x^{-3}$ and $\bar{\eta}$ denotes the anti-self-dual  't\,Hooft tensor. 
\section*{Gauge fields from superposition of merons or instantons}
The SU(2) gauge fields to be considered  are superpositions of instanton or meron fields
\begin{equation}
    \label{supo}
    A_{\mu}(x)= \sum_{i=1}^{N_P} h(i) a_{\mu}(x-z(i)) h^{-1}(i).
\end{equation}
The dynamical variables of these fields are  the positions of the centers
$z(i)$ of the instantons or merons and their
color orientations
\begin{equation}
    \label{color}
    h(i) = h_0(i) + {\rm i} {\bf h}(i) \cdot \mbox{\boldmath$\sigma$} \,, \qquad
    h_0^2(i)  + {\bf h}^2(i) = 1 .
\end{equation}
In general, the action of gauge fields generated  by superposition of either merons or instantons is logarithmically divergent in the infrared. By a judicious choice of the  color orientation,  finite values of the total action can be obtained.  This is achieved, e.g.,  by the following superposition  of  4 instanton or merons   
$$ A(x)= \sum_{i=1}^{3} R^{\pi}_i a(x-z(i)) +a(x-z(4)) \,, $$ 
with the color rotations $ R^{\pi}_i$  by $\pi$  around the three color directions $i$.   With this choice  
\begin{equation}
 x \to \infty: \quad A \sim \frac{1}{x^2}\,,\quad F[a] \sim \frac{1}{x^3}\,,\quad s(x)\sim \frac{1}{x^6}   \,,
\label{asy}
\end{equation}
the system is ``neutral'' and the action is finite. 
The constituents, instantons or merons,  are ``confined''. Removal of  a single constituent  gives rise to  an infinite action.  
Thus ensembles of instantons and merons exhibit similar infrared properties and in both cases  strong correlations between the constituents are required to guarantee a finite action.

The ensembles of  gauge fields (cf.\,Eq.(\ref{supo}))  to be constructed  in the following are defined by the partition function 
\begin{equation}
   \label{pathintegral}
Z = \int {\rm d} z_i {\rm d} h_i {\rm e}^{- \frac{1}{g^2} S[A(z_i,h_i)]         }\, ,
\end{equation}
with  the effective action $S$ identified with  the standard (Euclidean) action (cf.\,Eq.\,(\ref{acde}))
\begin{equation}
   \label{act0}
S=\int_V {\rm d}^4 x s(x)\, .
\end{equation}
This  definition of the partition function   guarantees that for finite $g^2$ a non-zero weight is assigned to neutral configurations only, i.e. to fields with the asymptotic behavior (\ref{asy}). 
After construction of   ensembles of field configurations  using the Metropolis algorithm,  vacuum expectation values of observables ${\cal O}$ can be computed 
\begin{equation}
   \label{pathintegral-obs}
\langle {\cal O} \rangle = \frac{1}{Z} \int {\rm d} z_i {\rm d} h_i {\rm e}^{- \frac{1}{g^2} S[A(z_i,h_i)]}
 {\cal O} [A(z_i,h_i)]  \,.
\end{equation}
In the numerical determination of these ensembles, the location $z(i)$ is restricted to
a hypercube
\begin{equation}
   -1\le z_{\mu}(i)\le 1 \,,\label{size}
\end{equation}
an equal number of instanton and anti-instantons (merons and antimerons) is used, and  in most of the applications the values of  size and coupling constant are chosen as 
 \begin{equation}
    \label{stco} 
\rho = 0.16\,, \quad g^2 =32\, . 
\end{equation}
\section*{Characteristics of instanton and meron ensembles}
\subsection*{Action densities}
\begin{figure}
\begin{center}
\includegraphics[width=.5\linewidth]{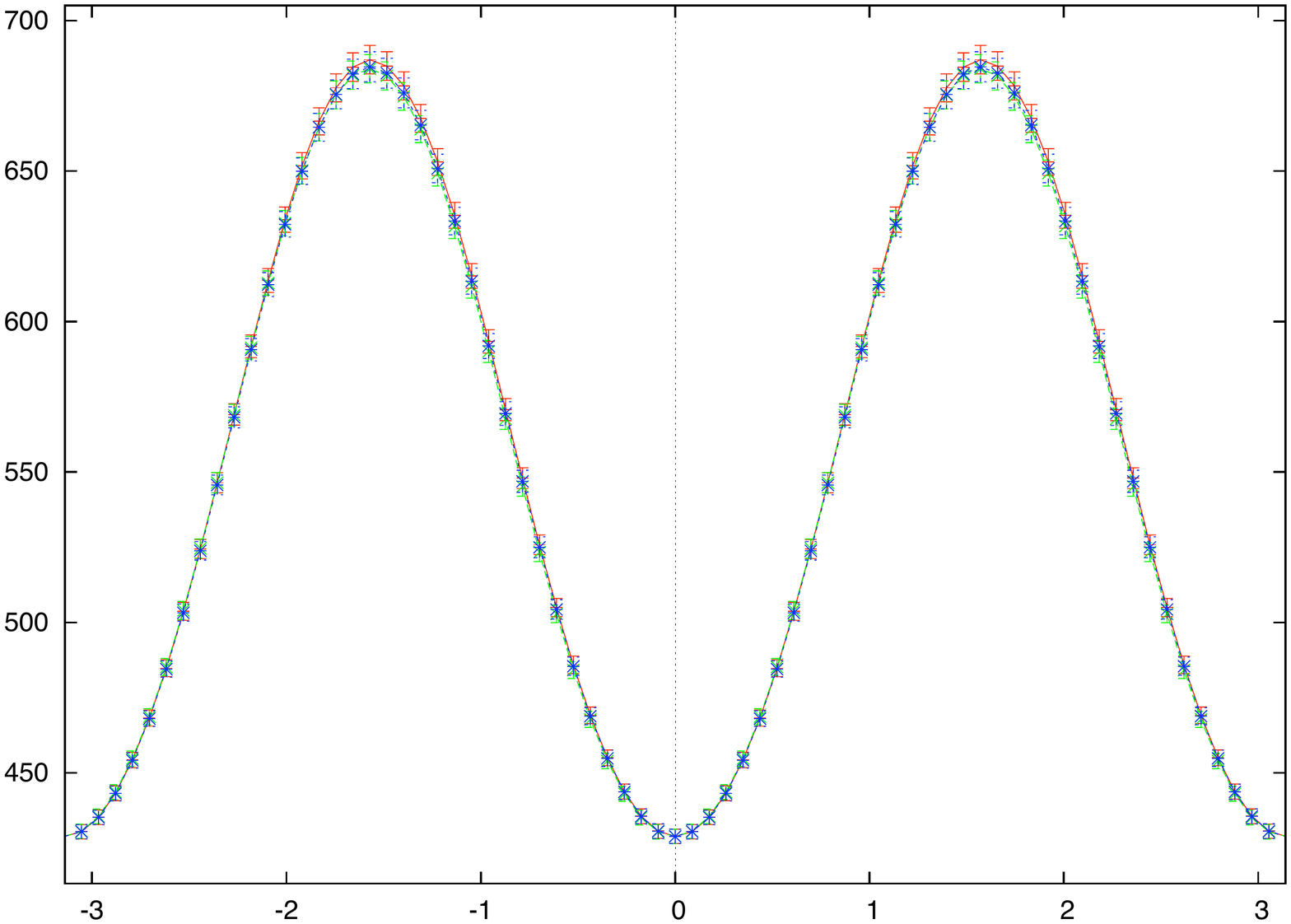}\includegraphics[width=.5\linewidth]{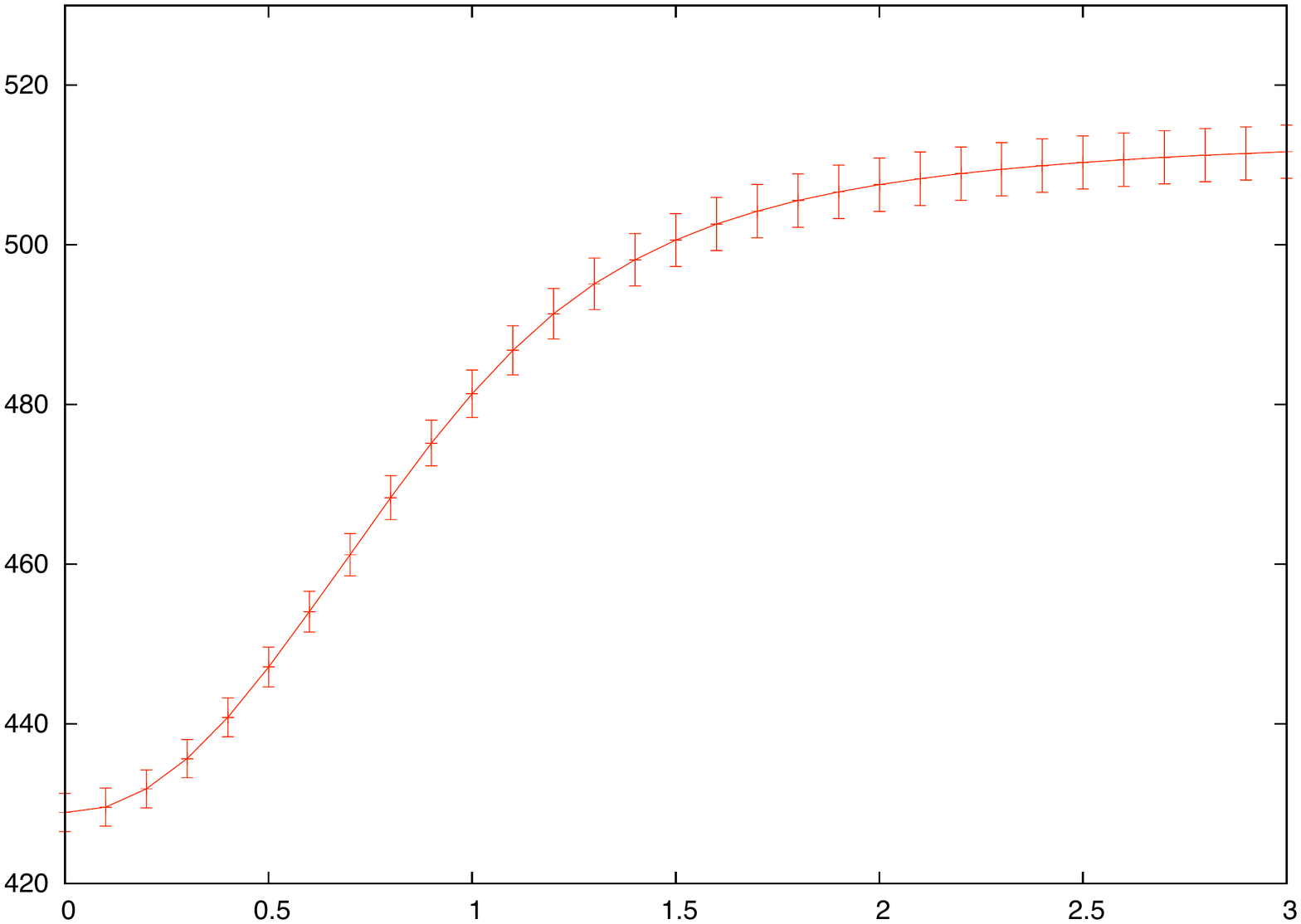}
\caption{Response of the action of 50 instantons to changes of the color orientation along three independent geodesics on the manifold $S^3$ of the color rotations (\ref{color})  (left) of the  instanton  closest  to the center  and to changes of  its position (right).}
\label{chac}
\end{center}
\end{figure}

In this section I will characterize qualitative properties of gauge fields in the ensembles defined by  (\ref{pathintegral}).
\begin{figure}
\vskip-.5cm
\includegraphics[width=1.\linewidth]{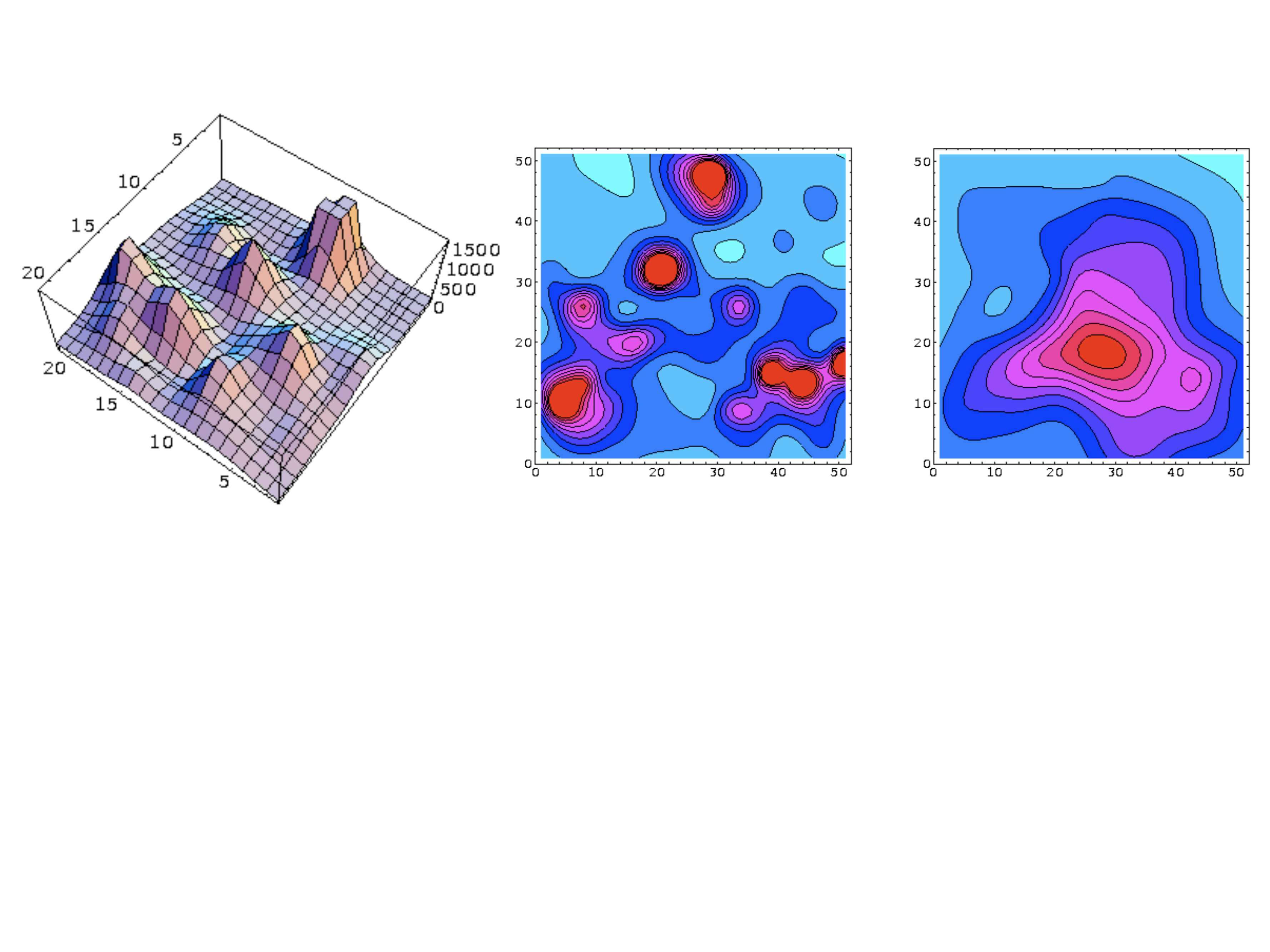}
\vskip -5cm
\caption{Action density along a plane for a gauge field of 500 merons (left) and lines of equal action density  ( $g^2=32$, middle) and  ($g^2 =\infty$ right). }
\label{meracd}
\end{figure}
In Fig.\,\ref{meracd} is shown a typical landscape of the action density displaying  peaks of single merons on top of a background. Background and single meron contributions to the total action are of the same order of magnitude. Unlike in the superposition of singular instantons, the appearance of single meron or instanton structures is not automatic, it rather  reflects the dynamics defined by the partition function (\ref{pathintegral}). This becomes evident in the  comparison (Fig.\,\ref{meracd})  with  the  action density  of a meron field of the corresponding stochastic ensemble ($g^2=\infty$). The values of the action density  of the two ensembles  
\begin{eqnarray}
g^2&=&32\,,\quad  \langle s \rangle = \;\;\;1500\,,\quad \; \; \;200\, <\,s(x) < \;11000\,,\nonumber\\
g^2&=&\infty\,,\quad  \langle s \rangle = 110 000\,,\quad 15 000 < s(x) < 330 000\,,
\label{cpacd}
\end{eqnarray} 
differ by almost two orders of magnitude. As a result of of the huge background  generated by the stochastic  superposition of the constituents  no single meron contribution ($s_{\text{max}} = 18 000$)   can be identified. The maximum close to the center of Fig.\ref{meracd} is a result of the stochastic superposition as  can be verified in an analytical calculation.  
The  reduction of the action from the strong coupling value is a first quantitative signature of the importance of  correlations  in the ($g^2\neq \infty$) ensembles. Another measure of the strength of the  correlations between constituents  is the response of  the action to changes of the color orientation or  of the position of a single meron or instanton \cite{LESZ}. As shown in Fig.\,\ref{chac}, under variations of  the color orientation of a single constituent  the total action is increased   by up to $60\%$ and by up to $20\%$ under variations of the position.  In a weakly interacting system of 50 instantons (instanton gas) one would expect a $2\%$ decrease of the action when removing a single instanton  and even smaller changes under variations of the color orientation. 
\subsection*{Confinement} 
In this section I  will  present  evidence that the phase of highly correlated  constituents is confining. 
In Yang Mills theories, the area law of  Wilson loops is the signature of confinement. The Wilson loop is defined by 
\begin{equation}
  \label{wilo}
  W = \frac{1}{2} \,\mbox{tr}\Big\{ P \exp {\rm i}g \oint_{\cal C}A_{\mu}(x){\rm d}x^\mu\Big\}\,,
\end{equation}
with the integral  ordered along the closed path ${\cal C}$. The following  results refer to rectangular paths with aspect ratio 2:1. 
For sufficiently large loops  the results for  the logarithm of the expectation value of the  Wilson loop can be  parametrized as a sum of constant, perimeter (${\cal P}$)
and area (${\cal A}$) terms,
\begin{equation}
  \label{fit}
  \ln  \langle W \rangle = \omega + \tau \, {\cal P} -\sigma {\cal A}\,,
\end{equation}
with $\sigma$ denoting the string tension.
The parameters of the fit  for ensembles of 500 merons ($\rho=0.16$) are  
\begin{equation}
  \label{stst500}
  \omega = -2.1,\; \tau = 2.3, \; \sigma = 22.9 
\end{equation}
for the stochastic ensemble ($g^2=\infty$). 
For ensembles with $g^2=32$ the following values  
\begin{equation}
  \label{stdy500}
  \omega = -0.7,\; \tau = 1.1, \; \sigma = 11.8 
\end{equation}
are obtained  if the  color orientations are dynamical variables while the  positions are randomly chosen but fixed, and 
\begin{equation}
  \label{dydy500}
  \omega = -0.7,\; \tau = 1.1, \; \sigma = 11.5
\end{equation}
for  the ensemble with dynamical color orientations and dynamical positions. 
\begin{figure}
\begin{center}
\includegraphics[width=.6\linewidth]{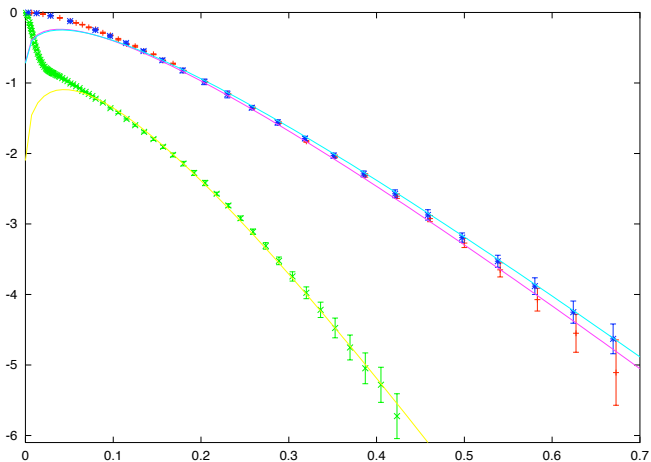}
\caption{Logarithm of a Wilson loop as a function of its area for the three  meron ensembles of Eqs.\,(\ref{stst500}) (green $\times$),  (\ref{stdy500}) (pink $-$), and (\ref{dydy500}) (blue $*$).}
\label{sco500sd}
\end{center}
\end{figure}
These three ensembles give rise to an area law  with  values of the  string 
tension differing by a factor of 2 between stochastic and dynamically correlated ensembles. The large difference in the dimensionless 
ratios formed by action density and string tension,  
\begin{equation}
  \label{si2s}
\frac{g^2 \langle s \rangle}{\sigma^2}\approx 500, \;\; (g^2=\infty)\,, \quad  \frac{g^2 \langle s \rangle}{\sigma^2}\approx 11 \; \;(g^2=32),  
\end{equation}
reflects the different dynamics of the ensembles. As in  lattice gauge theories, confinement   in the strong coupling limit  is a result of disorder of the system generated by the unconstrained fluctuations of the gauge fields. Suppression of the fluctuations if $g^2\neq \infty$ affects more strongly  the local action density than  the non local-Wilson loop and in turn leads to the strong decrease in the ratio $g^2 \langle s\rangle /\sigma$. 
Only minor changes of Wilson loops are obtained if, in addition to the color orientations,  also  the meron positions  are treated as dynamical variables. This result is in accordance  with the   above findings concerning the response to changes in the color orientation and position (Fig.\,\ref{chac}) of single  constituents. It suggests that the    confining phase is a  close relative of the nematic phase of liquid crystals with   strong correlations in internal  and  comparatively weak correlations in position space \cite{frle04}. 
\section*{Quantitative results}
\subsection*{Scaling and confinement}
Quantitative predictions for observables evaluated in the ensembles defined above require definition of a scale.   As in lattice gauge calculations the phenomenological value, $\sigma = 4.4\, \text{fm}^{-2}$, of the string tension is chosen to set  the scale.   
Rescaling  the different contributions to the Wilson loop (\ref{fit}), 
\begin{equation}
  \label{fitzins}
  \ln  \langle W \rangle = \omega + \tau \,\sqrt{\lambda}\, {\cal P} -\sigma \,\lambda\,{\cal A} ,
\end{equation}
\begin{figure}
\begin{center} 
\includegraphics[width=.65\linewidth]{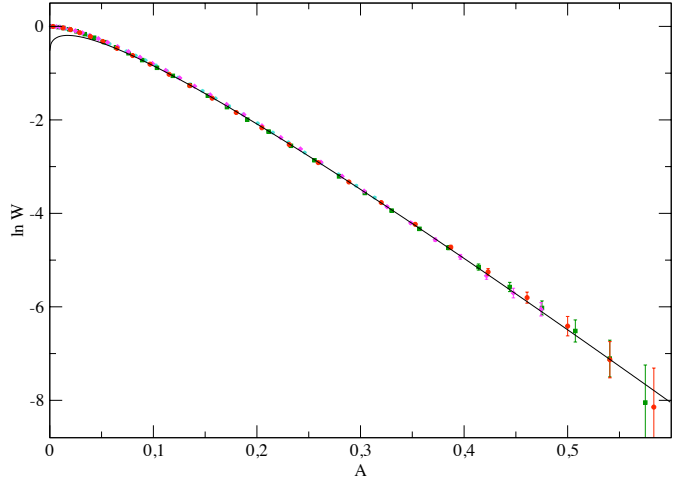}\vskip -.3cm
\caption{Logarithm of a Wilson loop as a function of its area for instanton ensembles  $N_I = 50$ (star), 100 (diamond), 200 
(square), 500 (circle). The area has been rescaled  with $\lambda$ given in Table \ref{inst1}.  Also shown is the  curve corresponding
 to the parametrization  (\ref{fitzins})   with the values of the parameters  given in (\ref{fitins}).}
\label{wiloins}
\end{center}
\end{figure}
and  fitting the scaling parameter $\lambda$ separately  for each ensemble gives rise to a universal curve as shown  in Fig.\,\ref{wiloins}. 
The values of the universal parameters and of the unit of length are for the instanton ensembles 
\begin{equation}
  \label{fitins}
\omega = -0.52,\quad \tau=1.18,\quad  \sigma = 19.0, \quad  1\, \mbox{u.l.}= 2.08  \sqrt{\lambda}\, \mbox{fm}\, .  
\end{equation} 
Similar results have been obtained for the meron ensembles.\\
As the Wilson loops, also other observables exhibit approximate independence of the number of degrees of freedom. Table\,\ref{inst1} summarizes for ensembles with $N_I$ instantons  the results for the vacuum expectation value of the action density $s$ and  the topological susceptibility (also in physical units), 
\begin{equation}
\chi = \Big(\frac{1}{ 8\pi^2}\Big)^{2}\int d^4 x \langle \tilde{s}(x)\tilde{s}(0)\rangle\,,
\label{susc}
\end{equation}
which is a measure of the strength of the fluctuations of the topological charge (\ref{toch}).  In comparison to the values  obtained from sum rules \cite{SHVZ79, NARI96} and from lattice gauge calculations  \cite{Campostrini:1983nr}
\begin{equation}
85\, \mbox{fm}^{-4} \le \langle s \rangle \le 260\, \mbox{fm}^{-4}\,,
\label{glco}
\end{equation}
the expectation value of the action density i.e. the gluon condensate is of the correct order of magnitude. It  receives about equal contributions  from  the background field generated by the superposition of the constituents and from the peaks of the fields of single constituents as suggested by Fig.\,\ref{meracd}. 
Also  the value of the topological susceptibility is in reasonable agreement with the lattice result $\chi^{1/4} \sim 215 \,$MeV \cite{Campostrini:1983nr}. 
The topological susceptibility is  dominated by the topological charge density of single constituents.   The single constituent contributions in the evaluation of the integral (\ref{susc}) yield the result,
$$\chi^{1/4}_I = 0.83\, n_I^{1/4}\,,\quad  \chi^{1/4}_M = 0.52 n_M^{1/4}\,,$$
which agrees  within 10$\,\%$ with the result of  the numerical evaluation (Table\,\ref{inst1}). Due to the dominance of single instanton properties similar values of this observable  in the liquid phase \cite{Schafer:1996wv} of singular gauge and in the confining phase of regular gauge instantons are obtained.  
\begin{table}
\begin{center}
{\begin{tabular}{||c|c|c||c|c|c|c|}  \hline\hline
$N_{I}$&$\langle s\rangle $&$\lambda\;\;$&$n_{I}$  &$\rho$& $\langle  s\rangle\;\;$&$\chi^{1/4}$ \\ \hline 
-&-&-&[fm$^{-4}]$ &[fm$^{1}$]&[fm$^{-4}]$  &[MeV]  \\ \hline \hline
500&5430  &1.0&$1.68 $ &0.33& $291 $ &162\\ \hline
200& 2490 &0.66&  1.54 &$ 0.27$ & $307 $ & 164\\ \hline
100&1350 &0.48&1.45& $ 0.23$ & $ 314$ &180\\ \hline
50 &651  &0.32&1.64& $ 0.19 $ & $340$& 190\\ \hline \hline
\end{tabular}}
\caption{Properties of instanton ensembles:  Vacuum expectation value of the action density $s$ defined in (\ref{acde}) and 
 the topological susceptibility $\chi$ (\ref{susc})  (also in physical units), 
 with the standard choice of the parameters (\ref{stco}) for ensembles of $N_I$ instantons  and with instanton  density $n_I= N_I/L^4$ and the values of the scaling parameter $\lambda$ (cf.\,(\ref{fitzins})) and the size parameter $\rho$ in physical units.}
\label{inst1}
\end{center} 
\end{table}
\\Table\,\ref{inst1}  shows that ensembles with increasing numbers of constituents ($N_I$) and located in a hypercube of   fixed volume (\ref{size}) describe, after rescaling,  systems of essentially the same density $n_I$ of instantons or merons. The increase in number is accounted for by the increase in volume of the hypercube. The value of the string tension determines the density of constituents.   Variations in the constituent size in the interval  $0.06\le\rho\le 0.25$  have been considered and found to  leave the scaling properties essentially unperturbed. Origin  of the  scaling properties  is  the scale independence  of the asymptotics of regular  gauge instantons or merons (Eq.\,\ref{mer}).  In ensembles of singular gauge instantons on the other hand,  $\rho$ controls  the strength of the asymptotics of  these fields (Eq.\,\ref{singin}) and therefore the overlap and in turn the   strength of the interaction of these constituents \cite{DIPP89}. 
\subsection*{Wilson loops and glueball masses}
\begin{figure}
\begin{center}
\includegraphics[width=.5\linewidth]{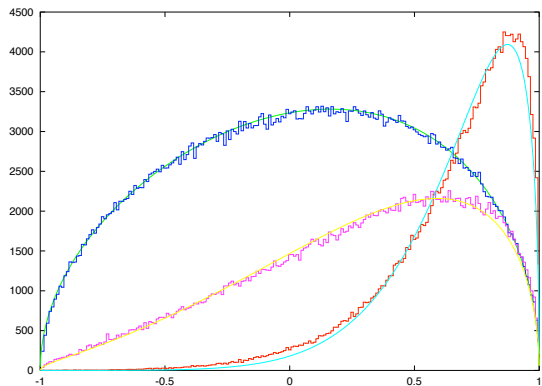}
\caption{Distribution of the values ($\cos \vartheta$) of Wilson loops  for an ensemble of configurations containing 500 merons with sizes  $0.48\times 0.24, 0.72\times 0.36, 1.\times 0.5 $ in comparison with the distributions (\ref{spre}).}
\label{diffwdis}
\end{center}
\end{figure}

In addition to the area law,  also  more detailed properties of Wilson loops have been found to agree with those of  lattice gauge calculations.  I mention  the Wilson loop distributions  shown in Fig.\,\ref{diffwdis} which, as similar   lattice results \cite{blnt05}, can be interpreted as distributions of a diffusion process on $S^3$, the group manifold of SU(2), given by  
\begin{equation}
  \label{spre}  
p(\cos \vartheta,t)= \frac{2}{\pi}\,\theta(t)\,\sum_{n=1}^{\infty}n\,\sin n\vartheta \,{\rm e}^{-(n^2-1)t} \,,
\end{equation}   
with  the diffusion  time $t$  determined by the expectation value of the Wilson loop.  Diffusion on $S^3$ entails ``Casimir scaling'' of the Wilson loop expectation values in higher representations which  has been observed in lattice gauge calculations \cite{bali00}. Also  in agreement with lattice gauge calculations is  the breakdown of Casimir scaling  observed for sufficiently large loops  for ensembles of gauge fields similar to the ones discussed here \cite{SZWA08}. 
\\In a study of  correlation functions of circular loops of equal radius $r$  which are parallel to each other and orthogonal to the direction of separation ($t$)
\begin{equation}
  \label{grog}
C_r(t)= \langle W_{r} ({\bf n}, {\bf x}_{0}, t)W_{r} ({\bf n}, {\bf x}_{0},0 )\rangle\, ,
\end{equation}
\begin{figure}[h]
\begin{center}
\includegraphics[width=.5\linewidth]{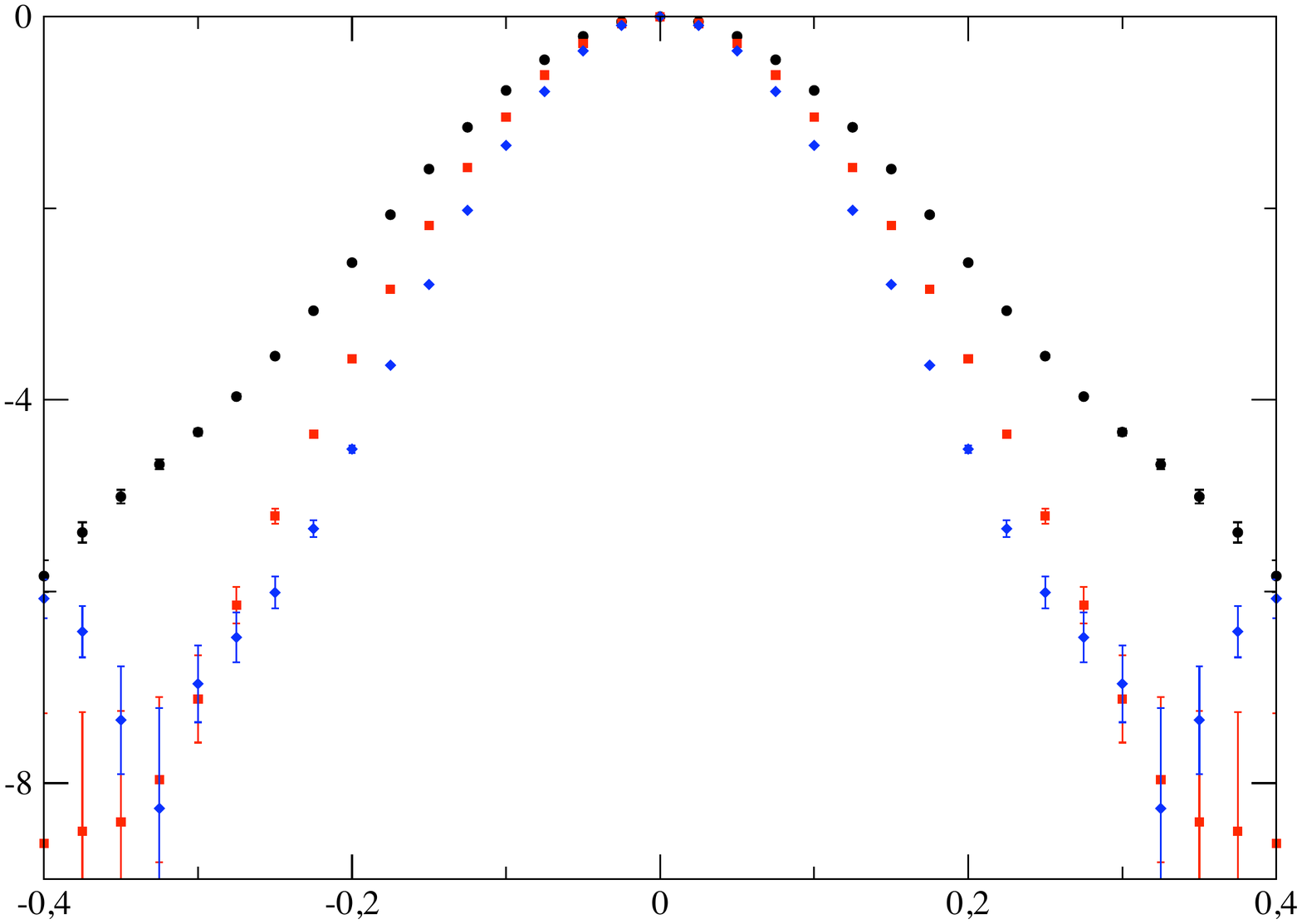}\includegraphics[width=.5\linewidth]{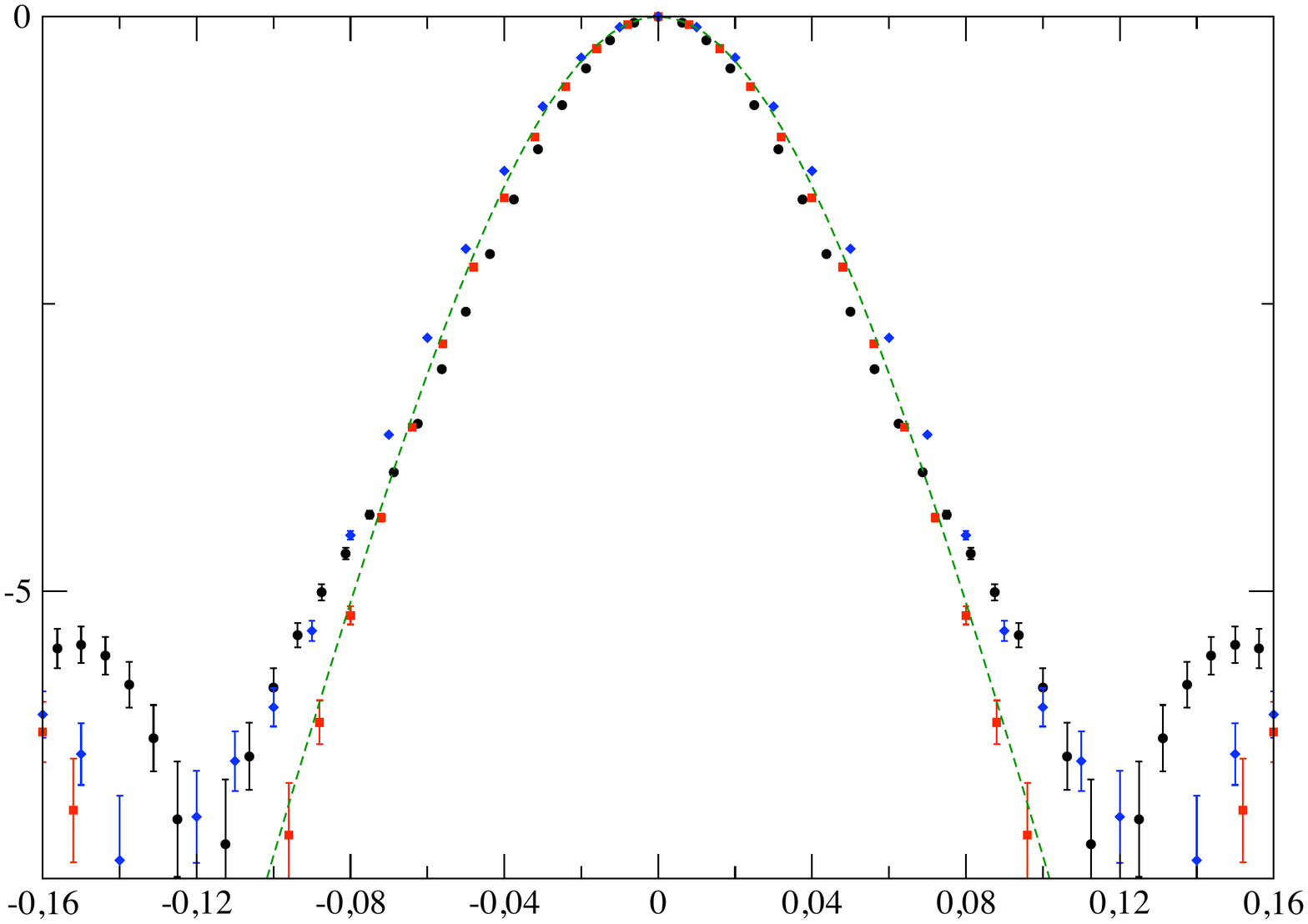}
\caption{Left: Logarithm of correlation functions of circular Wilson loops for an ensemble
 of 500 instantons  of size $\rho = 0.08$ as a function of separation in time $t$ for 3 values of the Wilson loop radius $r =0.25$ (black circles),
 $r=0.32$ (red squares), $r=0.40$ (blue diamonds) . \,  Right: The same as a function of the variable $rt$.}
\label{grogf}
\end{center}
\end{figure}
indications for formation of flux tubes  in connection with the area law have been found. If gauge strings are important degrees of freedom the application of  the Wilson loop operator to the
 vacuum should generate such a gauge string with energy  $E\approx  2\pi\sigma r$,  
and  the relevant variable for describing the correlation function should be $rt$.  Indeed  the three correlation functions are 
described approximatively by a universal curve in terms of this variable as shown in Fig.~\ref{grogf}.   A rough estimate of the string tension based on this calculation yields $ \sigma = (3.7  \pm  1.3) \,\text{fm}^{-2}\,.$ \\
As the last topic I discuss the glueball spectrum calculated via  (Euclidean) correlation functions 
\begin{equation*}
\langle {\cal O}(x) {\cal O}(0) \rangle  \sim \sum_n \langle \Omega | {\cal O}
| n \rangle {\rm e}^{-E_n x} \langle  n | {\cal O} | \Omega \rangle  \, .
\label{expsum}
\end{equation*}
\begin{figure}
\begin{center}
\includegraphics[width=.8\linewidth]{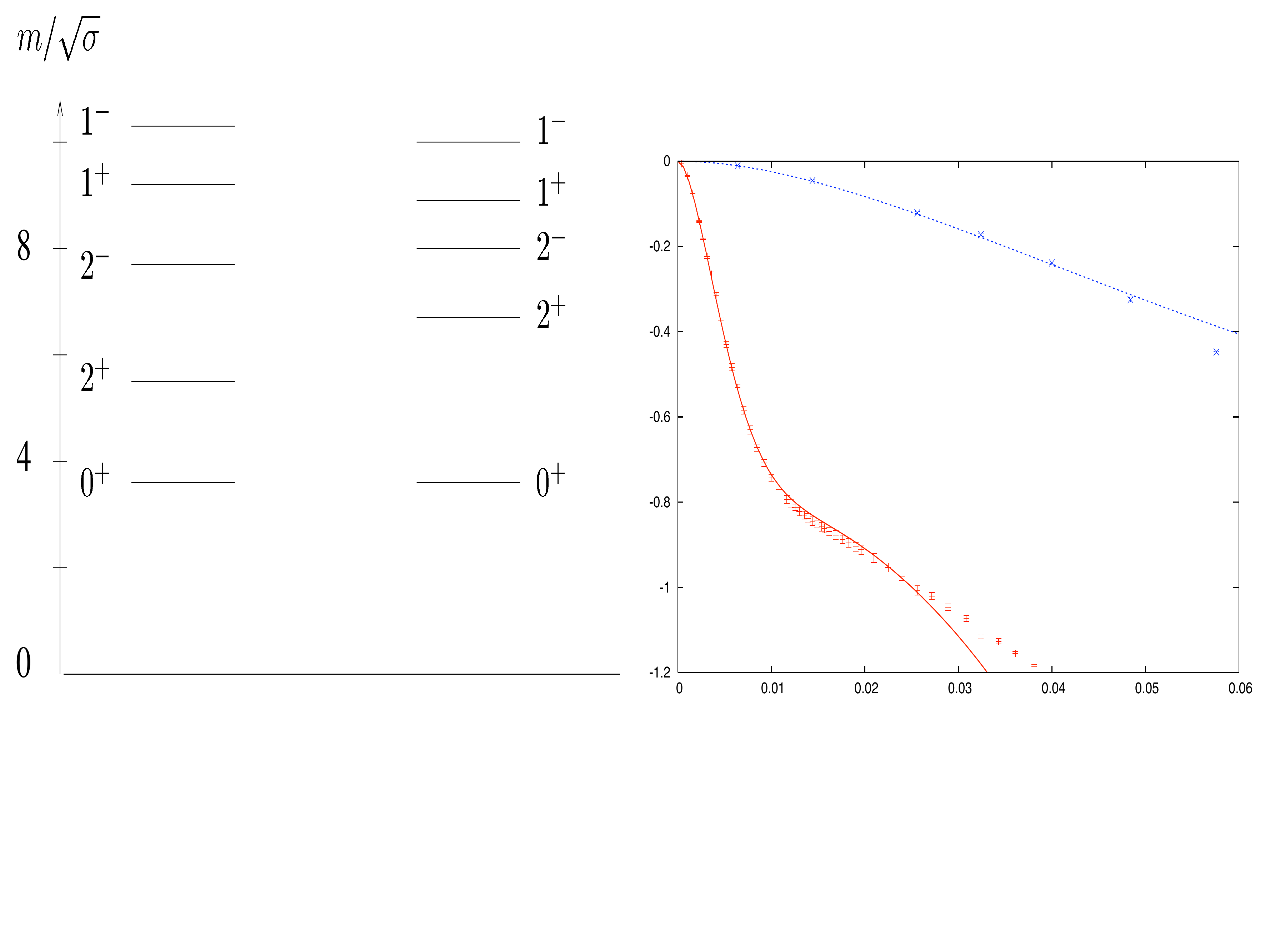}
\vskip -1.7cm
\caption{Left: Spectra of 
 glueball
states.  
 Lattice gauge theory \cite{tepe98} (left part) $\;$ Instanton   ensembles after rescaling (right part). Right: Enlarged portion of Fig.~\ref{sco500sd} for small size Wilson loops. Upper curve obtained for the ensemble  with dynamical meron positions and color orientations, lower curve for the stochastic ensemble ($g^2=\infty$). }
\label{spek}
\end{center}
\end{figure}
of appropriately chosen operators carrying the  quantum numbers of the glueballs. For sufficiently large separations,  the correlation function is dominated by the state of lowest energy which can be excited by the operator ${\cal O}$. For determination of the masses, local and non-local operators have been used and various ensembles differing in the size parameter and the number of constituents have been studied. The final result is shown in Fig.\,\ref{spek}. Qualitative agreement with respect to the ordering of the level has been reached. The scale however is not compatible with the scale set by the string tension. The values of the masses have been multiplied with a factor of 1.8 (1.5 for meron ensembles) to  reproduce the $0^+$ glueball mass  obtained in lattice gauge calculations. Determined by the requirement of the infrared finiteness of the action, the degrees of freedom  do not adequately account for the dynamics on the scale of the size of the glueballs (0.2 -0.5  fm)\cite{FOKE92}. Missing strength in the ultraviolet also shows up in the Wilson loops of small size as is shown in the right part of  Fig.\,\ref{spek}. While the stochastic ensemble with its unconstrained fluctuations generates a Coulomb-like behavior at small distances, the  ensembles of dynamical merons or instantons  fail in the description of Wilson loops in the perturbative regime. A remedy for the shortcoming in the calculation of the glueball spectrum could be the inclusion of  perturbative contributions in addition to the non-perturbative correlation functions, as applied   for  ensembles of singular gauge instantons \cite{SCSH95}. 
\section*{Conclusion}
The central achievement of these studies is the construction of ensembles of gauge fields  which  exhibit confinement. These fields are   obtained by superposition of regular gauge instantons or merons. It is remarkable that superposition of the gauge equivalent regular and singular instantons  gives rise to two distinctively different phases, the gas or liquid phase generated by singular  and the confining phase by regular gauge  instantons respectively.  The confining phase  owes its existence to the non-trivial requirement of finite action  in the presence of the slow asymptotic $1/x$ decay of the regular gauge instanton or meron fields.   Unlike  the disordered strong coupling limit,  the confining phase is an ordered phase similar to the nematic phase of liquid crystals. The  infrared  behavior of the constituents  is not only the source of confinement it's scale independence  also   induces the scaling properties of observables under changes in the number of constituents.
Besides various  properties of Wilson loops, such as the area law, Casimir scaling or flux tube formation, also 
 the action density (gluon condensate)  and  the topological susceptibility  are reasonably well reproduced.  These latter quantities are determined by the relative strength of the gauge fields close to the center of the constituents and of the background gauge field generated by the superposition of the constituents. If observables such as the topological susceptibility  are dominated by single instanton properties similar results in the liquid and in the confining phase can be expected.  Limits in the applicability of meron and instanton  ensembles  are encountered if short distance properties are important as is the case for the glueball spectrum and for small size Wilson loops. 
 
\end{document}